\documentclass[aps,twocolumn,prb,showpacs,floatfix,superscriptaddress]{revtex4}
\usepackage{graphics,amssymb,amsmath,epsfig}

\begin{document}

\title{Relaxation and Coarsening Dynamics in Superconducting Arrays}

\author{Gun Sang Jeon}
\affiliation{Center for Strongly Correlated Materials Research,
Seoul National University, Seoul 151-747, Korea}
\affiliation{Department of Physics and Institute for Basic Science
Research, Sung Kyun Kwan University, Suwon 440-746, Korea}

\author{Sung Jong Lee}
\affiliation{Department of Physics and Center for Smart Bio-Materials,
The University of Suwon, Kyunggi-do 445-743, Korea}

\author{M.Y. Choi}
\altaffiliation{Also at Korea Institute for Advanced Study, Seoul
130-012, Korea.}
\affiliation{Department of Physics, Seoul National University,
Seoul 151-747, Korea}

\begin{abstract}
We investigate the nonequilibrium coarsening dynamics in
two-dimensional overdamped superconducting arrays under {\em zero
external current}, where ohmic dissipation occurs on
junctions between superconducting islands through uniform resistance. 
The nonequilibrium relaxation of the unfrustrated array and also 
of the fully frustrated array, quenched to low temperature ordered states or
quasi-ordered ones, is dominated by characteristic 
features of coarsening processes via decay of point and line defects, 
respectively.  In the case of unfrustrated arrays, it is argued that 
due to finiteness of the friction constant for a vortex (in the 
limit of large spatial extent of the vortex), the typical length scale 
grows as $\ell_s \sim t^{1/2}$ accompanied by the number of point 
vortices decaying as $N_v \sim 1/t $.  
This is in contrast with the case that dominant dissipation occurs
between each island and the substrate, where the friction constant
diverges logarithmically and the length scale
exhibits diffusive growth with a logarithmic correction term. 
We perform extensive numerical simulations, to obtain results in reasonable agreement.
In the case of fully frustrated arrays, the domain growth
of Ising-like chiral order exhibits the low-temperature behavior
$\ell_q \sim t^{1/z_q}$, with the growth exponent $1/z_q$
apparently showing a strong temperature dependence in the
low-temperature limit.
\end{abstract}


\maketitle

\section{Introduction}

In a statistical system quenched from the random disordered state
to a low-temperature state below the transition temperature, the
average length scale $\ell$ of ordered domains typically grows in
time as a power law $\ell \sim t^{1/z}$, where the growth exponent
$1/z$ depends on the dimension of the space and of the relevant
order parameter, in addition to the conserved or nonconserved
nature of the latter in the relaxation dynamics.\cite{review}
Depending crucially on the dimension of the order parameter,
characteristic topological defects such as point vortices or
domain walls are generated in the initial disordered state and the
annihilation of these defects gives the main mechanism of
coarsening and phase ordering in the system.  The observed
self-similarity of these coarsening systems at different time
instants, is embodied in the so-called {\em dynamic scaling
hypothesis} of the equal-time spatial correlation function
of the order parameter.
%

One of the extensively studied cases is the system with the order
parameter dimension $n=2$ in two spatial dimensions $d=2$, i.e.,
the two-dimensional (2D) {\em XY} model, where the relevant
topological defect is the point
vortex.\cite{loft,mondello,bray,yurke} A convenient physical
realization of the 2D {\em XY} model can be found in the 2D
regular arrays of superconducting Josephson junctions, where the
phases of the superconducting order parameters correspond to the
$XY$ spins.
In such a superconducting array, frustration may be induced simply
by applying an external magnetic field.\cite{FXY}  It affects
thermodynamic properties in a crucial way, leading to a variety of
phase transitions.\cite{arrayreview}

In view of the remarkable diversity in the equilibrium properties,
also expected for the system are a variety of interesting dynamic
behaviors, particularly in relaxation toward equilibrium.
The quasi-long-range order present in the 2D pure $XY$ model,\cite{BKT}
describing the array without frustration (i.e., in the absence of 
a magnetic field), is expected to exhibit coarsening through 
annihilation of
vortex-antivortex pairs, giving rise to characteristic algebraic
relaxation.  The vortex dynamics and the resulting scaling behavior
in the 2D $XY$ model has been an issue for some time,
concerning late-time scaling and true nature of the asymptotic growth law. 
It is also of interest to examine the effects of the additional
long-range order on the relaxation in the fully frustrated array
(with half the flux quantum per plaquette),\cite{FXY,FFXY} where
the relaxation dynamics is expected to be dominated by coarsening
processes accompanying decay of domain walls and also point
defects; the latter corresponds to the corners on the domain
walls.

The main purpose of this paper is to probe the properties of the
nonequilibrium relaxation dynamics associated with coarsening processes
in the unfrustrated and fully frustrated superconducting arrays
under rapid quenching from the high-temperature disordered state to
the low-temperature ordered or quasi-ordered states.
We would like to understand what kind of specific features emerge
in the relaxation dynamics of these systems governed by
the resistively-shunted-junction (RSJ) dynamic equations,
especially in connection with the coarsening processes.
The time-dependent behaviors of various one-time and two-time
correlations involving energy and order parameters such as vortex density
and chirality are investigated numerically.

In the case of an unfrustrated array, we argue that the friction constant 
of a slow moving vortex remains finite in the limit of large extent of 
the vortex;\cite{korshu_mob} this leads (via a simple force-velocity
relation) to a power-law growth of length scales with growth exponent $1/2$, 
together with the excess number of vortices decaying as
$\tilde{N}_v
\sim t^{-\alpha_v}$ with $\alpha_v =1$.
Extensive simulations indeed show that $\alpha_v$ takes a value
that is very close to unity in a wide range of temperatures.  This
is in contrast to the simulation results of the ordinary phase 
ordering process based on either Monte Carlo algorithms 
or {\em simple Langevin dynamics} of the $XY$ model (see section III for details), 
where a logarithmic correction factor is observed in the time dependence of 
the growing length scale.  The latter case corresponds to the situation in 
Josephson-junction arrays where dissipation occurs between each island and 
the substrate. 

In relation to the coarsening dynamics of fully frustrated arrays, 
Lee, {\it et al.} \cite{sjl_ffxy} performed dynamic simulations on 
the fully frustrated $XY$ models, based on the simple Langevin 
equations for \textsl{XY} phase angles.  It was shown that the domain growth
exponents are temperature-dependent and that there exist two regimes of
domain-growth morphology.  Later, coarsening was also studied in the dual 
representation, i.e., the lattice Coulomb gas of fractional vortices, 
through the use of Monte Carlo methods on the vortex degrees of freedom.\cite{f2_cg}  
The latter work, even though based on different dynamics, confirmed most features 
of the previous results and gave some numerical evidence for unbinding of the so-called
single-step kinks at a finite temperature. 
Those rich features of coarsening in the fully frustrated $XY$ model
can be partly attributed to the existence of the Ising-like chiral
degrees of freedom in addition to the underlying phase degrees of freedom. 
Here, in this work, we perform dynamic simulations on the coarsening of 
the fully frustrated Josephson-junction arrays, governed by the RSJ equations, 
and examine whether temperature-dependent coarsening also emerges. 
It is found that, similarly to the previous works, the domain growth 
exponents are strongly temperature-dependent: The exponents are proportional 
to the temperature in the low-temperature limit and increase monotonically 
with the temperature. 

\section{Equations of Motion}

We begin with the set of equations of motion for the phases
$\{\phi_i\}$ of the superconducting order parameters in an
$L\times L$ square array.  In the RSJ
model with the fluctuating twist boundary conditions,\cite{FTBC}
they read:
\begin{equation} \label{eq:dyn1}
{\sum_j}' \left[
\frac{d \widetilde{\phi}_{ij}}{dt}
+ \sin (\widetilde{\phi}_{ij}-{\bf r}_{ij} \cdot {\bf \Delta})
 + \zeta_{ij}\right]=0 ,
\end{equation}
where we have employed the abbreviations
$\widetilde{\phi}_{ij}\equiv \phi_i{-}\phi_j{-}A_{ij}$ and
${\bf r}_{ij}\equiv {\bf r}_i{-} {\bf r}_j$, and the primed summation
runs over the nearest neighbors of grain $i$. The position of
grain $i$ is represented by ${\bf r}_i = (x_i , y_i )$ with the
lattice constant set equal to unity while the gauge field $A_{ij}$
is given by the line integral of the vector potential ${\bf A}$:
\begin{equation}
A_{ij} \equiv \frac{2\pi}{\Phi_0} \int_{{\bf r}_i}^{{\bf r}_j}
{\bf A}\cdot d{\bf l}
\end{equation}
with the flux quantum $\Phi_0 \equiv hc/2e$. The frustration
parameter $f$, which measures the number of flux quanta per
plaquette, is given by the directional sum of the gauge field
$A_{ij}$ around a plaquette:
\begin{equation} \label{eq:sumA}
f \equiv {1 \over 2\pi}\sum_P A_{ij} .
\end{equation}
In Eq. (\ref{eq:dyn1}) the energy and the time have been expressed
in units of $\hbar I_c / 2e$ and $\hbar/2eRI_c$, respectively,
with single-junction critical current $I_c$ and shunt resistance
$R$. The thermal noise current $\zeta_{ij}$ is assumed to be
white, satisfying
\begin{equation}
\langle \zeta_{ij}(t{+}\tau) \zeta_{kl} (t) \rangle = 2 k_B T
\delta (\tau) (\delta_{ik}\delta_{jl} {-} \delta_{il}\delta_{jk})
\end{equation}
at temperature $T$.  Henceforth we set the Boltzmann constant $k_B
\equiv 1$.  The dynamics of the twist variables ${\bf \Delta}
\equiv (\Delta_x, \Delta_y )$ is governed by
\begin{equation} \label{eq:dyn2}
\frac{d\Delta_{a}}{dt} =\frac{1}{L^2} \sum_{\langle ij
\rangle_{a}} \sin(\widetilde{\phi}_{ij}-\Delta_{a}) + \zeta_{a} ,
\end{equation}
where $\sum_{\langle ij \rangle_{a}}$ denotes the summation over
all nearest-neighboring pairs in the $a$-direction $(a = x, y)$
and $\zeta_{a}$ satisfies
\begin{equation}
\langle \zeta_{a}(t{+}\tau) \zeta_{b} (t) \rangle = {2 T \over
L^2} \delta_{ab} \delta (\tau).
\end{equation}
Note that in equilibrium the above set of equations of motion
leads to a Gibbs distribution with the Hamiltonian for the
frustrated \textsl{XY} model
\begin{equation} \label{XY}
H = -\sum_{\langle i,j\rangle} \cos(\tilde{\phi}_{ij} -{\bf
r}_{ij}\cdot\Delta ),
\end{equation}
where the fluctuating twist boundary conditions have been
incorporated.
%
In numerical simulations, we have integrated directly the
equations of motions in Eqs.~(\ref{eq:dyn1}) and (\ref{eq:dyn2})
via the modified Euler method with time steps $\Delta t = 0.05$.
Simulations have been performed on square arrays with linear size
$L = 400$ (unfrustrated arrays) and $L = 128$ (fully frustrated
arrays).

To study the relaxation and coarsening in the superconducting arrays,
we let the systems evolve from random initial configurations at given
temperatures and measure the following quantities:
\begin{enumerate}
\item The excess amount of topological defects such as vortices and domain walls
\begin{equation}
\tilde{N}(t) \equiv N(t) - N(\infty) ,
\end{equation}
where $N(t)$ is the total amount (number or length) at time $t$ 
and $N(\infty)$ refers to the amount in equilibrium.  [In the case of 
vortices $N$ denotes the sum of the total number of vortices (plus charges) and 
that of antivortices (minus charges).]  It is
understood that the appropriate ensemble average over random
initial configurations is to be taken.
\item Excess energy relaxation defined to be
\begin{equation}
\tilde{E}(t) \equiv E (t) - E(\infty)
\end{equation}
where the energy (per site) $E(t)$ at time $t$ is given by
\begin{equation}
E(t) \equiv - { 1 \over {L^2}} \sum_{\langle i,j \rangle}\left\langle
\cos \left[ \widetilde{\phi}_{ij}(t)-{\bf r}_{ij} \cdot {\bf \Delta}(t)\right]
\right\rangle
\end{equation}
with the summation over all nearest-neighboring pairs and $\langle
\cdots \rangle$ denoting the ensemble average over random initial
configurations.
\item The equal-time spatial correlation function
\begin{equation}
C(r,t)= \langle O^{\ast}({\bf r}, t)\, O(0, t)\rangle
\end{equation}
of the order parameter $O({\bf r}, t) \equiv e^{i\phi ({\bf r}, t)}$.
\end{enumerate}
According to the dynamic scaling hypothesis, the correlation
function assumes the 2D scaling form
\begin{equation} \label{eq:dynamic_scaling}
C(r,t)= r^{-\eta} g(r/\ell )
\end{equation}
with the appropriate scaling function $g(x)$, where $\ell =\ell
(t)$ is the typical length scale of ordered domains (at time $t$).
While the exponent $\eta$ vanishes ($\eta =0$) in the usual
non-critical quenching to long-range ordered states, it takes
nonzero values in the case of quenching to critical states,
depending on the temperature $T$, i.e., $\eta=\eta(T)$.

\section{Coarsening and vortex dynamics in unfrustrated arrays}

It is well known that the unfrustrated system ($A_{ij}=0$),
described by the \textsl{XY} model in equilibrium, is critical
below the Berezinskii-Kosterlitz-Thouless transition temperature
$T_c \ (\approx 0.89)$.\cite{bkt_weber}  Previous works on relaxation 
of the $XY$ model dealt mostly with the time-dependent 
Ginzburg-Landau equations (either the soft-spin version or the hard-spin version). 
The hard-spin version, where the magnitude of each spin is fixed, is described by 
the {\em simple} Langevin equations for phases:
\begin{equation} \label{eq:xy_lang}
 \frac{d \phi_i }{dt} = - { {\delta H } \over \delta \phi_i }+ \zeta_{i}.
\end{equation}
where $ H = -\sum_{\langle i,j\rangle} \cos(\phi_{i} - \phi_{j}) $
is the Hamiltonian and the noise $\zeta_{i}$ satisfies 
$ \langle \zeta_{i}(t{+}\tau) \zeta_{j} (t) \rangle = {2T}\delta_{ij} \delta (\tau)$. 
Experimentally, this {\em simple} Langevin dynamics corresponds to the situation 
where dissipation through the resistance between superconducting islands 
and the substrate dominates over those through the junction resistance.

In this model with no frustration, it was argued that the friction 
constant $\gamma$ of a slow moving vortex depends logarithmically on 
the size (spatial extent) $r$ of the vortex, i.e., $\gamma \sim \ln(r/r_0)$, 
where the cutoff length scale $r_0$ may
be taken as the size of the vortex core.\cite{ryskin} Combining
this result with the coulomb force law of $F(r) \sim -r^{-1}$, we
get the relation $\dot{r} \ln(r/r_0) \sim -r^{-1}$, which leads to
the conclusion that the time dependence of the vortex number
exhibits logarithmic correction to the power law: $N_v \sim
t^{-1}\ln t$. The typical length scale, e.g., the separation
between vortices, is then expected to follow \cite{korshu_mob,ryskin} 
 $\ell_s \sim (t/\ln t)^{1/2}$.
Numerical confirmation of this prediction
has turned out rather tricky and subtle.\cite{rojas}

Here, we apply a similar argument to the coupled RSJ dynamics of
the superconducting arrays.  Interestingly, the mobility of a
vortex whose motion is governed by the RSJ dynamics is found not
to vanish in the large-separation limit ($r \rightarrow \infty$)
but to remain finite.  This can be easily seen from the following
argument based on the equations of motion in Eq.~(\ref{eq:dyn1})
for the phase variables.  Equation~(\ref{eq:dyn1}) can be
rewritten in the following form \cite{shenoy}
\begin{equation} \label{eq:dyn_new}
\sum_j M_{ij} {{d \phi_j } \over dt } = - { {\delta H } \over \delta \phi_i }
 - {\sum_j}' \zeta_{ij},
\end{equation}
where $-M_{ij}$ is the discrete Laplacian matrix and the
Hamiltonian $H$ is given by Eq.~(\ref{XY}).  In order to obtain
the mobility of an isolated vortex, we compute the energy
dissipation for an isolated vortex moving with a constant velocity
$v$ along, say, the $x$ direction.  The corresponding phase
configuration takes the form $\phi_v (x{-}vt, y)$, where $\phi_v
(x, y)$ represents the phase at site $i \equiv (x, y)$ in the
presence of an isolated (static) vortex.  At low temperatures,
where the noise term may be neglected, we use
Eq.~(\ref{eq:dyn_new}) and write the energy dissipation in the
form \cite{ryskin}
\begin{equation} \label{eq:dissipation}
{{dE} \over {dt} } = \sum_i { {\delta H } \over \delta \phi_i }{{d
\phi_i } \over dt } = - \sum_{ij} M_{ij} {{d \phi_j } \over dt }
{{d \phi_i } \over dt }.
\end{equation}
Inserting the vortex solution $\phi_v (x{-}vt, y)$ in the above
equation and working in the continuum notation, we get the
following expression for the energy dissipation
\begin{equation}
{{dE} \over {dt} } = - \gamma_v v^2,
\end{equation}
where the friction constant (or the inverse of the mobility)
$\gamma_v $ is given by
\begin{equation} \label{eq:friction}
 \gamma_v \approx -\int dx dy \partial_x \phi_v (x{-}vt,y) \nabla^2
 \partial_x \phi_v (x{-}vt,y).
\end{equation}
In the derivation of Eq.~(\ref{eq:friction}) from Eq.~(\ref{eq:dissipation}), 
it has been noted that in the continuum limit, the discrete Laplacian matrix $-M_{ij}$
turns into $\nabla^2$ and the time derivative becomes velocity multiplied by 
the spatial derivative with respect to $x$ for the ansatz vortex configuration.  
In the limit of zero vortex velocity ($v \rightarrow 0$), one can
easily see from dimensional argument that the mobility becomes
finite for a vortex of an infinite extent.  Namely, the above
integral for $\gamma_v$ is convergent in the large-size limit,
owing to the presence of the additional Laplacian derivative term
$\nabla^2$.  Since we have $\partial_x \phi_v (x,y) \sim r^{-1}$
for a single isolated vortex, the integrand in the above equation
goes as $\sim r^{-4}$, which is infrared convergent in two
dimensions. 

\begin{figure}
\centerline{\epsfig{file=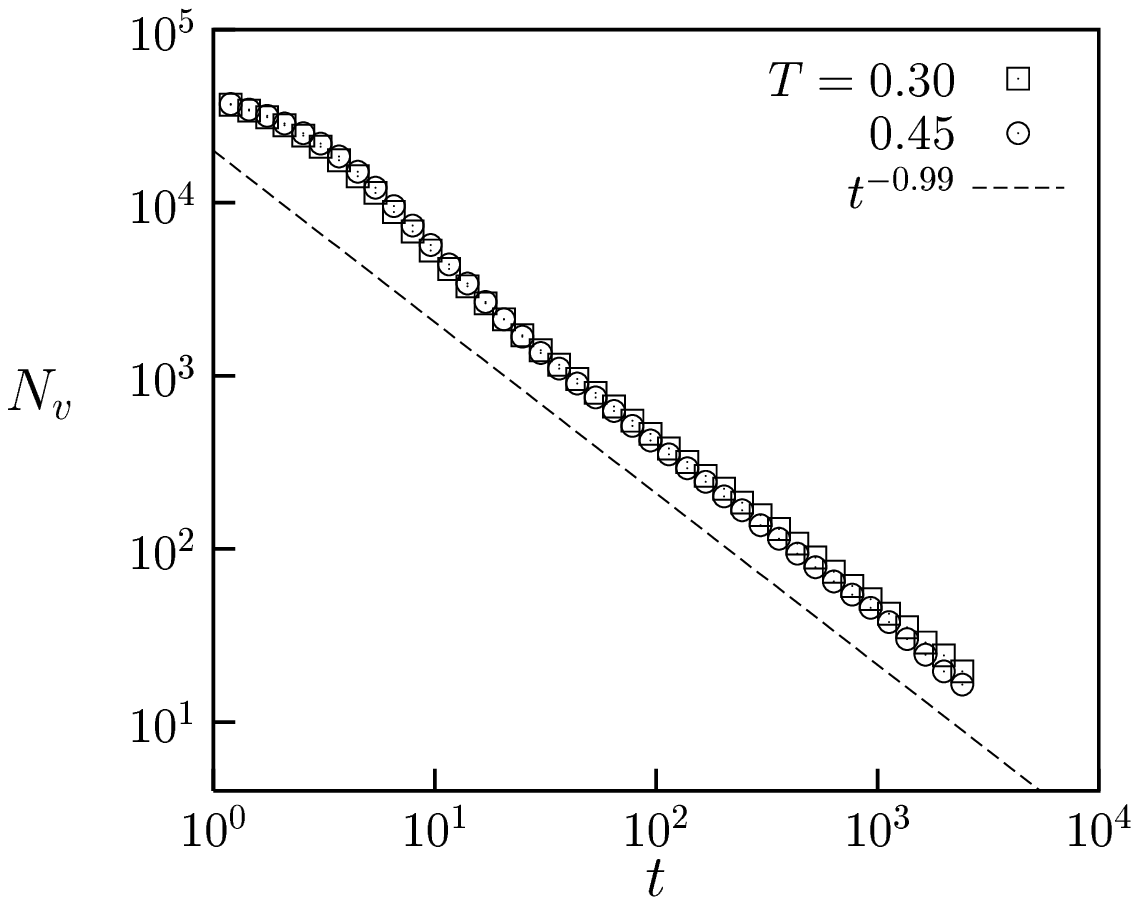,width=7.4cm}} \centerline{(a)}
\vspace*{0.5cm} \centerline{\epsfig{file=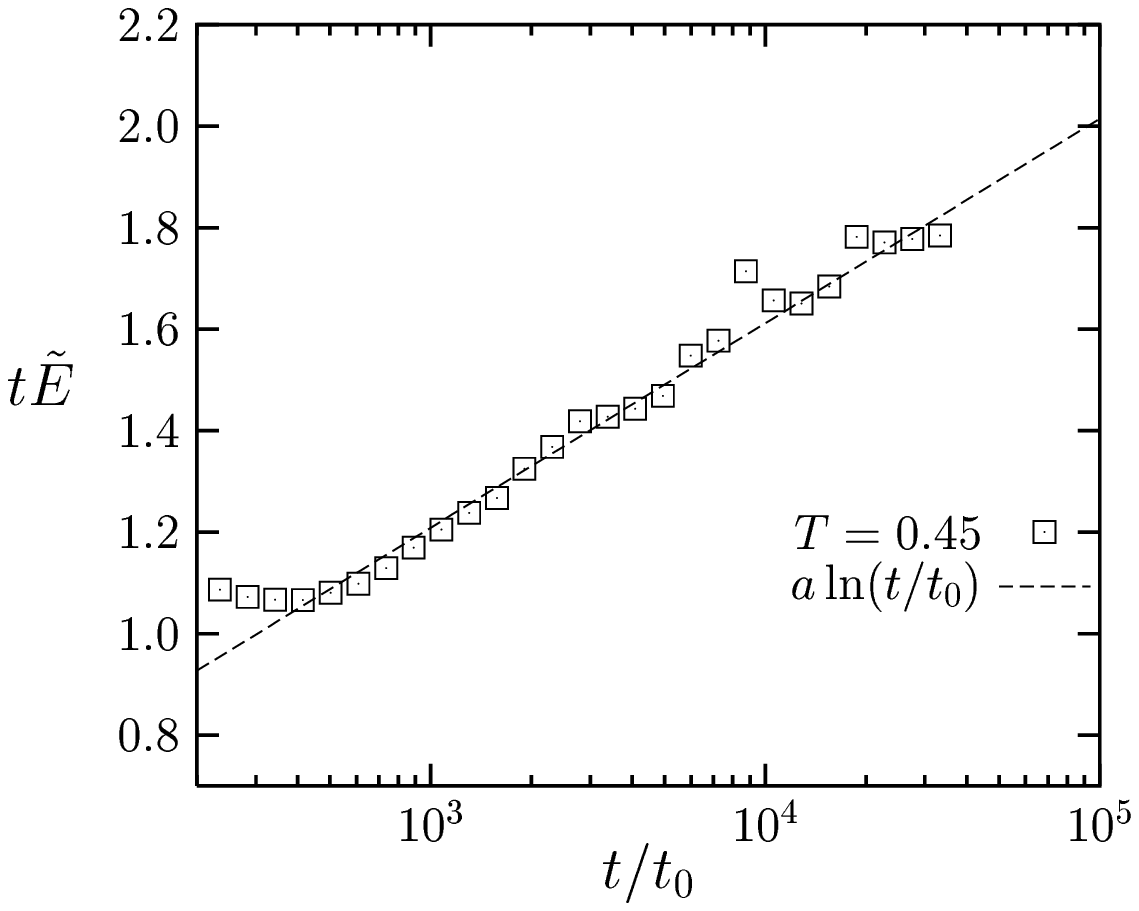,width=7.4cm}}
\centerline{(b)} \caption{Relaxation of (a) the vortex number
$N_v$ at temperature $T = 0.30$ and $0.45$ and of (b) the excess
energy $\tilde{E}$ at $T=0.45$ in the unfrustrated array of size
$L=400$.
In (a) the power-law behavior $t^{-0.99}$ is also plotted whereas
(b) displays the best fit to the form $\tilde{E} = a t^{-1}\ln(t/t_0)$ 
with $a = 0.175$ and $t_0 = 0.06$.
}
\label{fig:xy}
\end{figure}

Since the friction constant converges quickly to a finite value
$\gamma_0$ as the extent of the vortex is increased, the equation
of motion resulting from the force balance
reads
\begin{equation}
\gamma_0 \dot{r} \sim - r^{-1},
\end{equation}
from which one can estimate the annihilation time for a
vortex-antivortex pair of size $\xi$ to be $t \sim (\xi / \xi_0
)^2$, with the cutoff length scale $\xi_0$ corresponding to the
size of the vortex core.  This can be interpreted as the growth
law for the typical (average) separation between vortices as $\xi
\sim t^{1/2}$, which in turn leads to the vortex number $N_v \sim
\xi^{-2} \sim t^{-1}$.  Figure~\ref{fig:xy}(a) shows the
relaxation of the number of vortices at temperatures $T=0.3$ and
$0.45$.  Observed in the late-time regime are good fits with $N_v
\sim t^{-0.99}$ and $t^{-0.98}$ at $T=0.45$ and at $T=0.3$,
respectively, in excellent agreement with the expected $t^{-1}$
behavior.

This result may be used to obtain the energy relaxation behavior:
On the basis of the dynamic scaling hypothesis and 
the Porod law,\cite{review} which refers to the power-law singularity
in the short (scaled) distance limit of the equal-time spatial correlation
function due to the presence of topological defects, 
it is known that the excess energy for the $O(n)$ model with
$n=2$ scales as \cite{review}
\begin{equation}
\tilde{E}  \sim  \ell^{-2} \ln \left({\ell\over r_0}\right),
\end{equation}
where the typical length scale may be taken to be separation
between vortices ($\ell \approx \xi$) and the cutoff to be the
vortex core size ($r_0 \approx \xi_0$) in our system.  Noting the
simulation result $\xi \sim t^{1/2}$ presented above, we get 
$\tilde{E} \sim t^{-1} \ln (t/t_0)$ with $t_0$ being approximately
the initial microscopic time scale taken by a vortex to move a
single lattice spacing.  The excess energy versus time measured at
$T=0.45$ is shown in Fig.~\ref{fig:xy}(b) and indeed fitted quite
well to the expected form $\tilde{E} \approx at^{-1} \ln (t/t_0)$
with $a = 0.17$ and $t_0 = 0.06$.  We have also tried power-law
fitting of the form $\tilde{E} \sim t^{-\alpha}$, to obtain the
effective exponent $\alpha \approx 0.89$.  Such a value which is
smaller than unity can be understood as coming from the
logarithmic factor.

\begin{figure}
\centerline{\epsfig{file=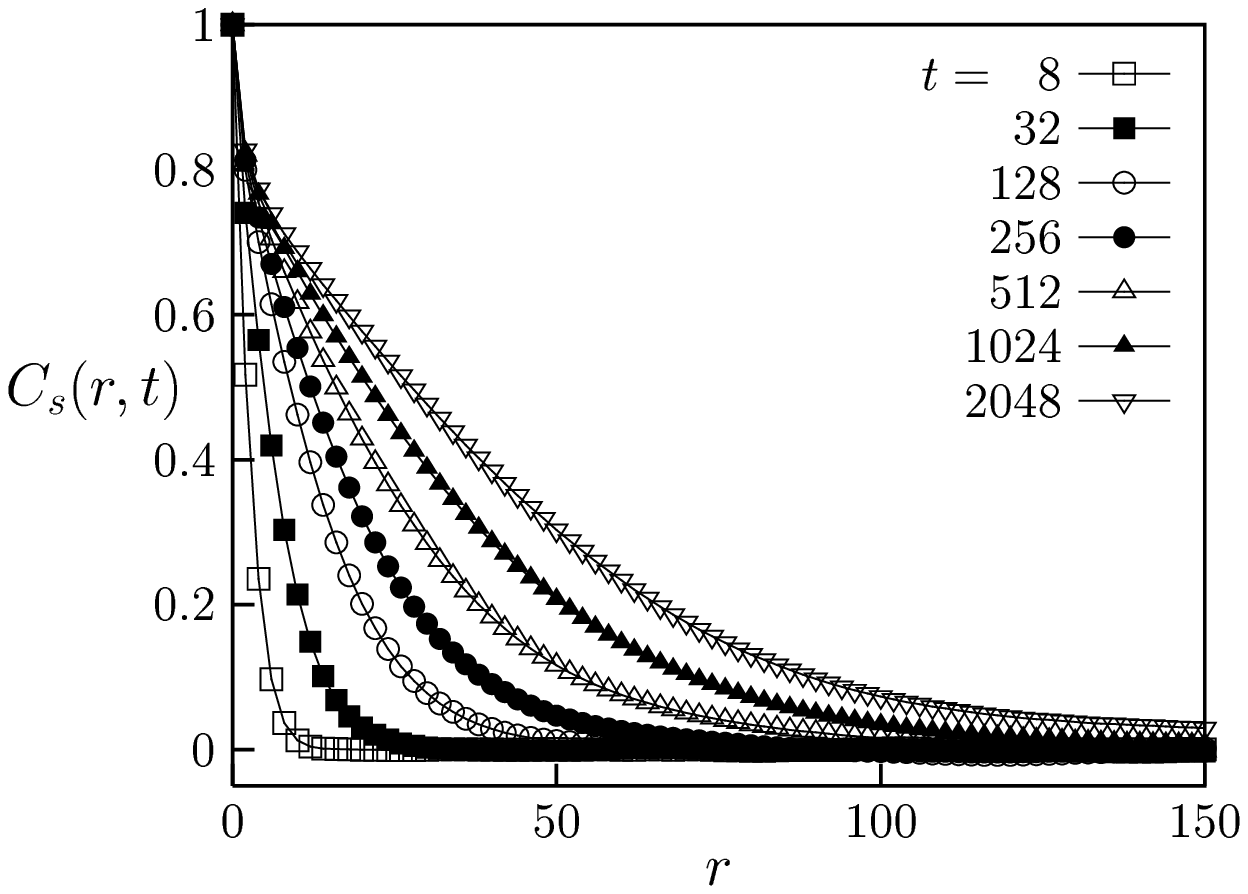,width=7.4cm}} \centerline{(a)}
\vspace*{0.5cm} \centerline{\epsfig{file=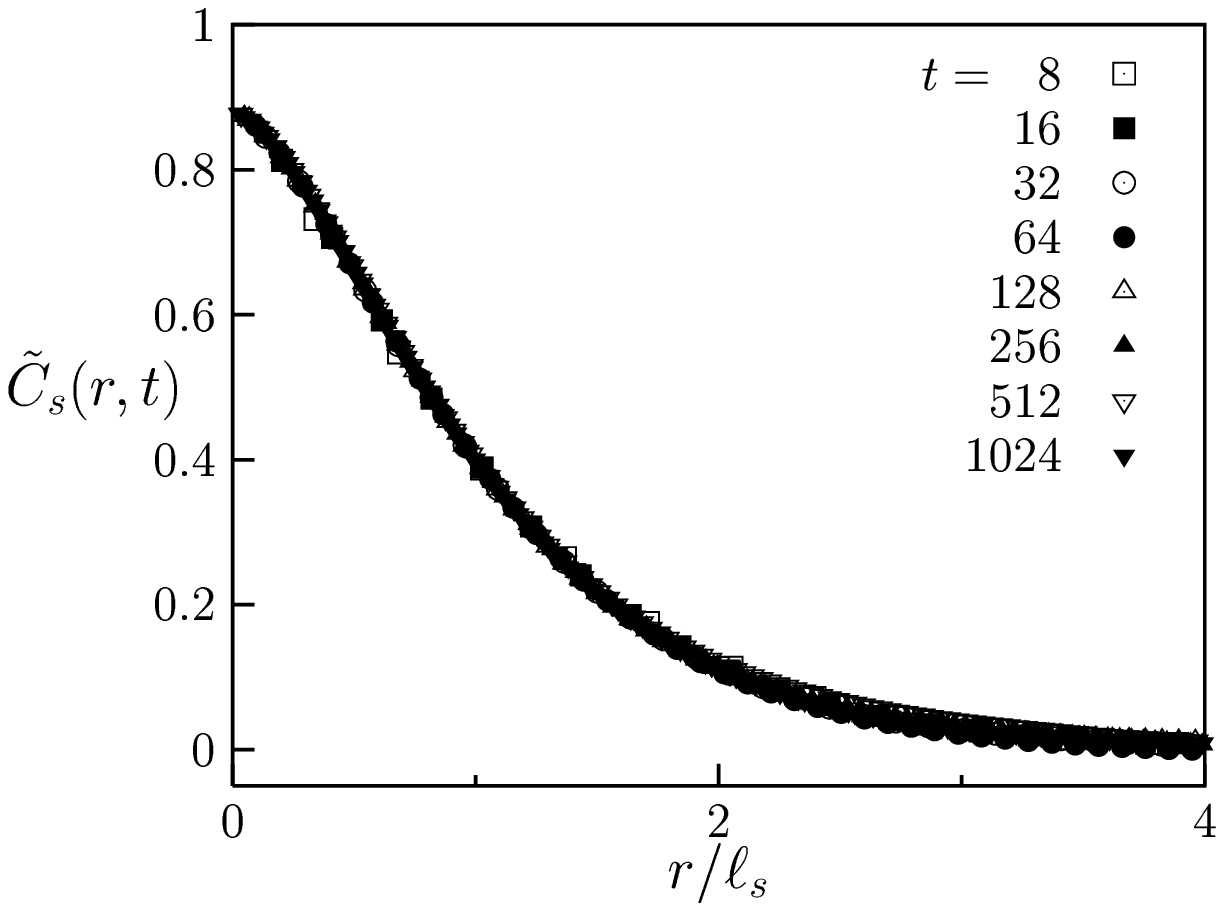,width=7.4cm}}
\centerline{(b)} \vspace*{0.5cm}
\centerline{\epsfig{file=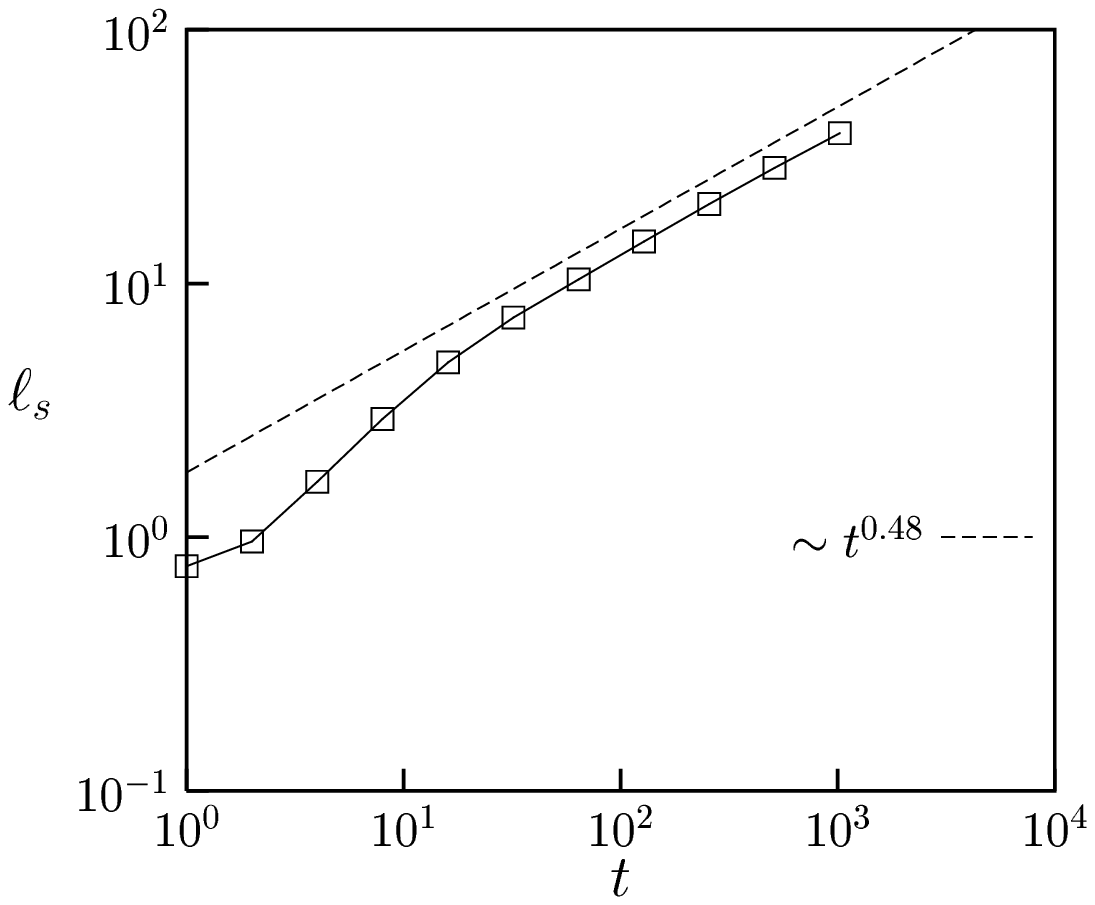,width=7.4cm}} \centerline{(c)}
\caption{(a) Spatial behavior of the equal-time phase correlation
function at various time stages in the unfrustrated array of size
$L = 400$ at temperature $T=0.45$. (b) Scaling collapse of the
data in (a), plotted in terms of the rescaled correlation function
with $\eta = 0.083$ and the scaling length $\ell_s$ defined in the
text. (c) Scaling length versus time at $T=0.45$, exhibiting
power-law growth $\ell_s \sim t^{0.48}$. } \label{fig:xy_coarse}
\end{figure}

Another test on the growing length scale in the unfrustrated array
can be performed by calculating the equal-time phase correlation
function
\begin{equation} \label{eq:spin_eqtc}
C_s (r, t) \equiv {1 \over {L^2}} \sum_{i} \left\langle \cos
\left[\phi_{i+{\bf r}}(t)- \phi_{i}(t)\right] \right\rangle .
\end{equation}
Figure~\ref{fig:xy_coarse} shows the equal-time spatial
correlation function of phases at temperature $T=0.45$.  Behavior
of the phase correlation function $C_s (r, t)$ with distance $r$
is displayed in Fig~\ref{fig:xy_coarse} at various time stages $t
= 8$ to $2048$.  Critical dynamic scaling has been attempted with
various values of $\eta$:
For each trial value of $\eta$, the rescaled function $\tilde{C}_s
(r,t) \equiv r^{\eta} C_s (r,t)$ has been plotted in terms of the
rescaled distance $r/\ell_s$, where the length scale $\ell_s$ as a
function of time $t$ has been obtained from the condition
$\tilde{C}_s (r{=}\ell_s, t) = C_0 \equiv 0.4$.  The value of
$\eta$ yielding the best scaling collapse has been chosen, which
is shown in Fig.~\ref{fig:xy_coarse}(b).  The resulting optimal
value $\eta = 0.083$ is in fair agreement with the value from the
theoretically predicted form of the equilibrium correlation
exponent \cite{villain} 
$\eta_{eq} \approx T/ 2\pi + T^2 / 4 \pi \approx 0.0877$.
The length scale $\ell_s$ obtained in this way is presented in
Fig.~\ref{fig:xy_coarse}(c), where one finds that $\ell_s \sim
t^{1/z_s}$ with $1/z_s \approx 0.48$, slightly below the expected
value $1/2$.  We have also performed simulations at various
temperatures, and found that the value of $1/z_s$ is nearly
constant, showing only small variations in the range $1/z_s
\approx 0.47$ to $0.48$.  The reason behind this small but
apparently systematic deviation is not clear at this stage.
Presumably, it is related to the fact that we are working with a
lattice system where small local barriers exist for the motion of
point vortices.  At the early-time stage when the vortices are
close together, these barriers are negligible compared with
inter-vortex forces; at the late-time stage when vortices are far
away, local energy barriers are likely to hinder appreciably the
vortex motion, possibly resulting in smaller values of the
effective growth exponent.

\section{Coarsening in the relaxation dynamics of fully frustrated arrays}

\begin{figure*}
\parbox{0.45\textwidth}{
\centerline{\epsfig{file=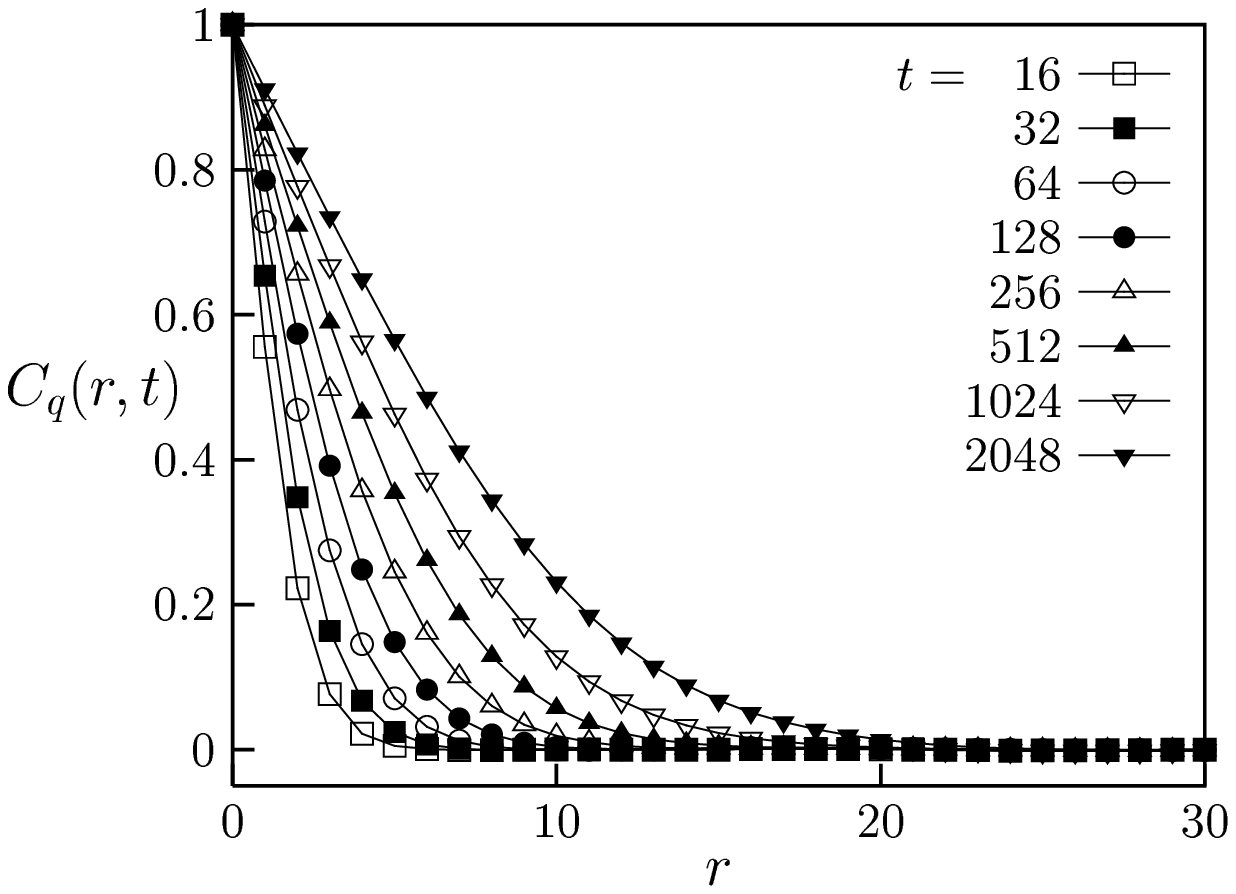,width=8cm}} \centerline{(a)}
\vspace*{0.5cm} \centerline{\epsfig{file=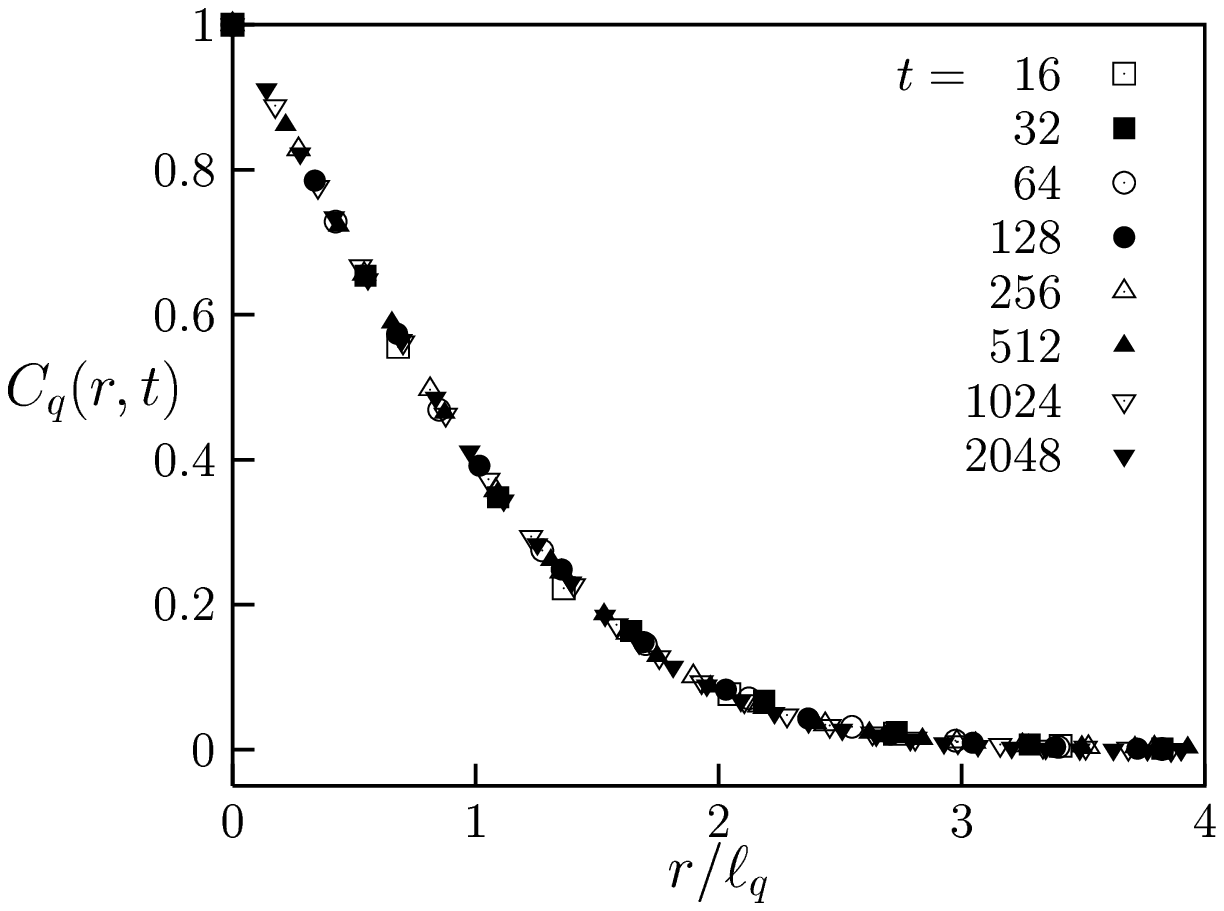,width=8cm}}
\centerline{(b)} } \hspace{0.05\textwidth}
\parbox{0.45\textwidth}{
\centerline{\epsfig{file=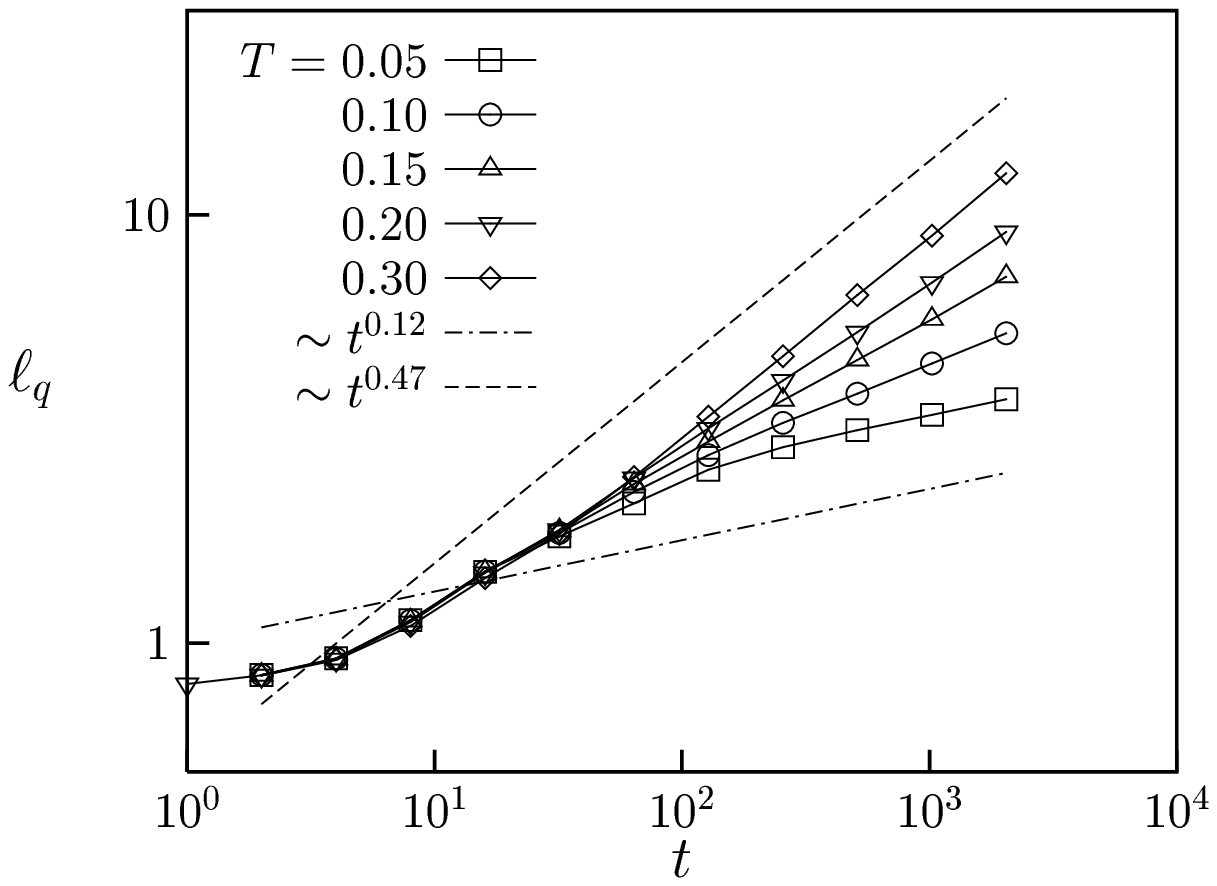,width=8cm}} \centerline{(c)}
\vspace*{0.5cm} \centerline{\epsfig{file=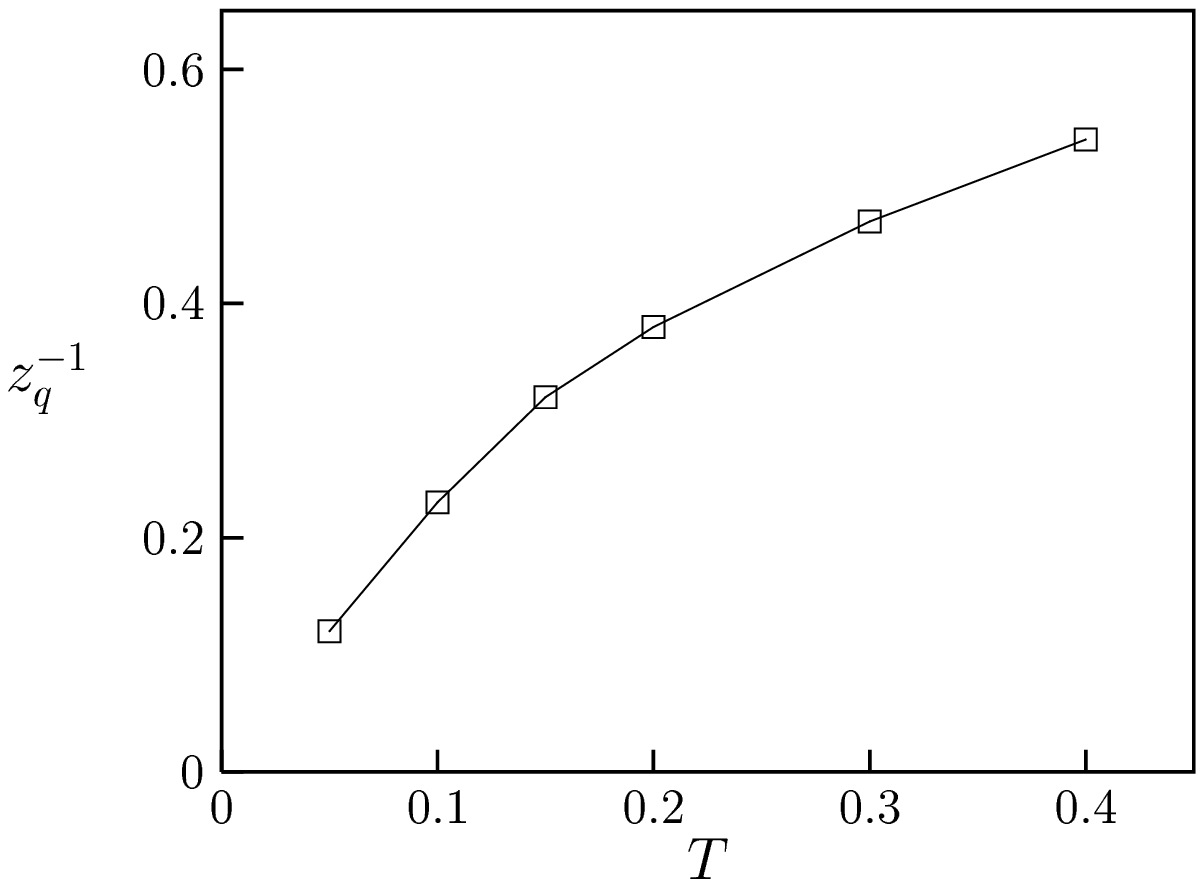,width=8cm}}
\centerline{(d)} } \caption{ (a) Spatial behavior of the
equal-time chirality correlation function at various time stages
in the fully frustrated array of size $L = 128$ at temperature
$T=0.15$.  (b) Scaling collapse of the data in (a) with the
appropriate scaling length $\ell_q$.  (c) Scaling length versus
time at various temperatures $T= 0.3,\, 0.2,\, 0.15,\, 0.1,\,
0.05$ from above, revealing temperature-dependent growth.  (d)
Growth exponent versus temperature, manifesting that growth
becomes slower as the temperature is lowered.}
\label{fig:ffxy_coarse}
\end{figure*}

We now consider the fully frustrated array ($f{=}1/2$) with the gauge field:
\begin{equation}
A_{ij} =
\left\{
\begin{array}{cl}
0 & \hbox{ for } {\bf r}_j = {\bf r}_i + \hat{\bf x} \\
\pi x_i & \hbox{ for } {\bf r}_j = {\bf r}_i + \hat{\bf y},
\end{array}
\right.
\end{equation}
which corresponds to the system of $XY$ spins with one
antiferromagnetic coupling per plaquette. The fully frustrated
array has the discrete Z$_2$ symmetry in addition to the
underlying U(1) symmetry, yielding the antiferromagnetic
long-range order in the chirality at temperatures $T<T_c
\,(\approx 0.45)$.\cite{FFXY} Due to the existence of the
chirality order parameter, the coarsening dynamics in a fully
frustrated array proceeds in a manner quite different from that of
the unfrustrated array.

In order to probe nonequilibrium coarsening of a fully frustrated
array, we measure the equal-time spatial correlations of the phase
and the chirality.  The phase correlation function in the fully
frustrated array can be defined in the same way as
Eq.~(\ref{eq:spin_eqtc}) for the unfrustrated array, only if it is
assumed that the correlations are measured only for even lattice
spacings along both the axes.  Restricting the position vectors to
those with components of even lattice spacings corresponds to
forming a superlattice, each plaquette of which consists of four
plaquettes of the original lattice.  Such a superlattice system
yields a ferromagnetic ground state within itself at zero
temperature.
Meanwhile, the chirality correlation function is defined as
follows:
\begin{equation}
C_q (r,t) \equiv {1\over N^2}\sum_{\bf R} (-1)^{x+y}\langle
q_{{\bf R}+{\bf r}} (t) q_{\bf R}(t)\rangle ,
\end{equation}
where the chirality
\begin{equation}
q_{\bf R}(t) \equiv {\rm sgn} \sum_P \sin
\left[\widetilde{\phi}_{ij}(t) - {\bf r}_{ij}\cdot {\bf \Delta}(t)
\right] ,
\end{equation}
with $\sum_P$ denoting the directional plaquette summation of
links around dual lattice site $\bf R$, describes the degrees of
freedom associated with the Z$_2$ symmetry.  The factor
$(-1)^{x+y}$, in the definition of $C_q (r, t)$, where ${\bf r}
\equiv (x, y)$, takes care of the {\em staggered} nature of the
chirality ordering.  Due to the existence of these two kinds of
degrees of freedom, the relaxation and ordering process proceeds
via annihilation of two kinds of defects i.e., point defects
(tending to restore the U(1) symmetry) and line defects (domain
walls tending to restore the chiral $Z_2$ symmetry). Point defects
correspond to the corners of line defects where fractional charges
of magnitude $1/4$ reside.\cite{korshu} Hence the annihilation
processes of these two kinds of defects are closely coupled with
each other. 
Note that conventional point vortices with integer charges, possessing
high energies, are annihilated very quickly in the early stage of coarsening; 
thus they are irrelevant in the late-time coarsening behavior of the system. 

Here we present our simulation results for the ordering dynamics
of only the chirality since the phase ordering exhibits similar
features except for the additional $\eta$ exponent due to the
criticality.  Figure~\ref{fig:ffxy_coarse} shows the equal-time
correlation function $C_q (r, t)$ of the (staggered) chirality at
temperature $T=0.15$.  Spatial behavior of $C_q (r, t)$ is
exhibited in Fig.~\ref{fig:ffxy_coarse}(a) at various time stages
$t=16$ to $2048$.  Scaling of the correlations can be obtained
similarly to the case of the unfrustrated array, except for the
obvious fact that $\eta_q =0$, reflecting the emergence of true
long-range order.  The resulting nice scaling collapse is shown in
Fig.~\ref{fig:ffxy_coarse}(b).  The scaling length $\ell_q$ is
plotted versus time in Fig.~\ref{fig:ffxy_coarse}(c) at various
temperatures, from which one obtains the relation $\ell_q \sim
t^{1/z_q}$ with the appropriate growth exponent $1/z_q$.

In contrast to the case of the unfrustrated array, the growth 
exponent in the fully frustrated array depends
strongly on the temperature, as shown in
Fig.~\ref{fig:ffxy_coarse}(d). For example, we have the value
$1/z_q \approx 0.38$ at temperature $T=0.2$ but it decreases to
$1/z_q \approx 0.32$ at $T=0.15$. At low temperatures, in
particular, $1/z_q$ depends almost linearly on the temperature,
which has also been observed in the ordering dynamics of the fully
frustrated {\em XY} model based on the simple Langevin 
equation. \cite{sjl_ffxy} 
%
On the other hand, the temperature dependence of the domain growth
emerges only after some time ($t \gtrsim t_q \approx 30$); before
that time ($t \lesssim t_q$) the domain growth apparently does not
change with the temperature. Such temperature independence at the
early-time stage is a distinctive feature that is not found in the
simple Langevin dynamics. This indicates that dissipation at 
the early-time stage is dominated by a mechanism that is different 
from the temperature-dependent activation dynamics operating at the
late-time stage. 

\begin{figure}
\parbox{0.22\textwidth}{
\centerline{\epsfig{file=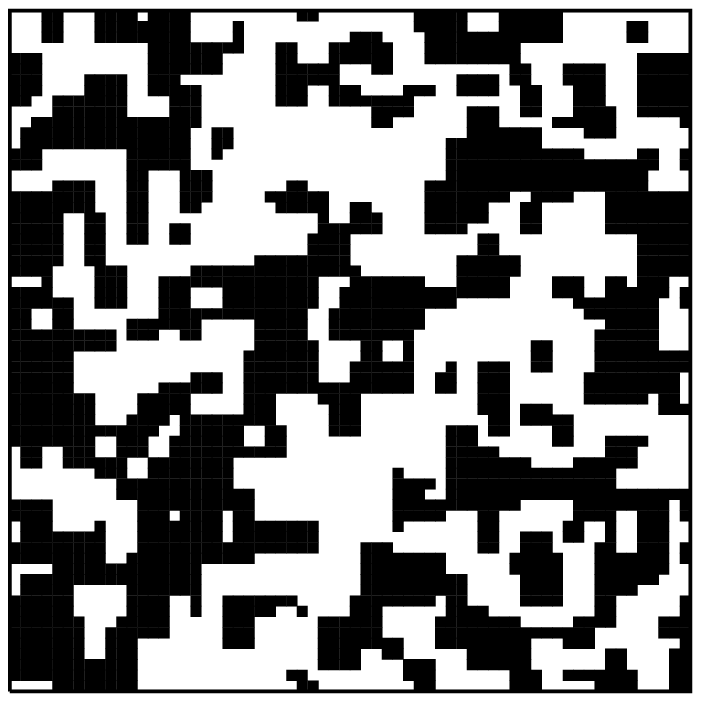,width=3.5cm}}
\centerline{(a)}
\vspace*{0.5cm}
\centerline{\epsfig{file=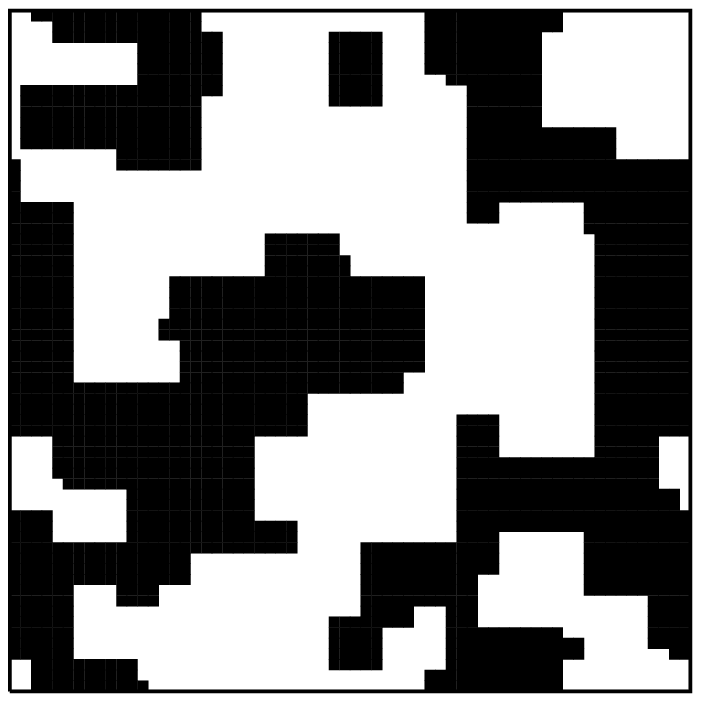,width=3.5cm}}
\centerline{(b)}
}
\hspace{0.02\textwidth}
\parbox{0.22\textwidth}{
\centerline{\epsfig{file=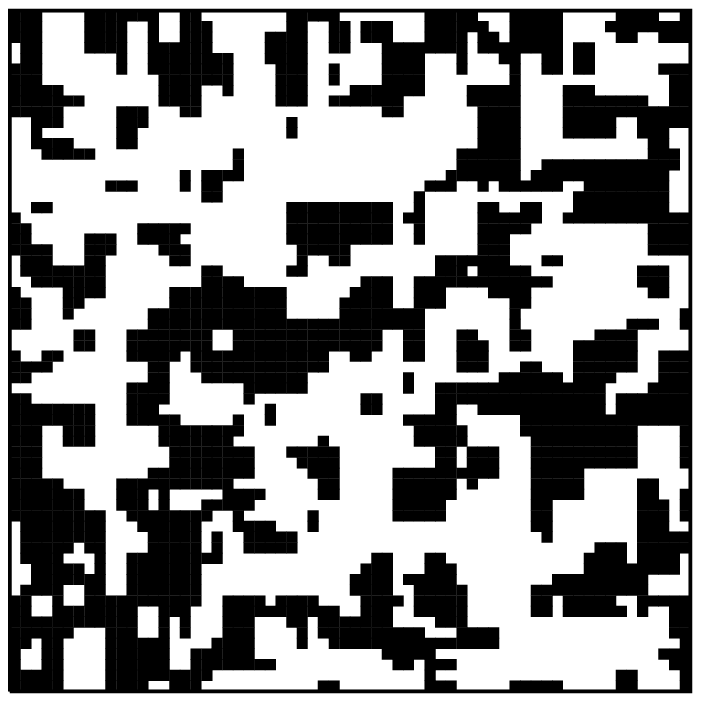,width=3.5cm}} \centerline{(c)}
\vspace*{0.5cm} \centerline{\epsfig{file=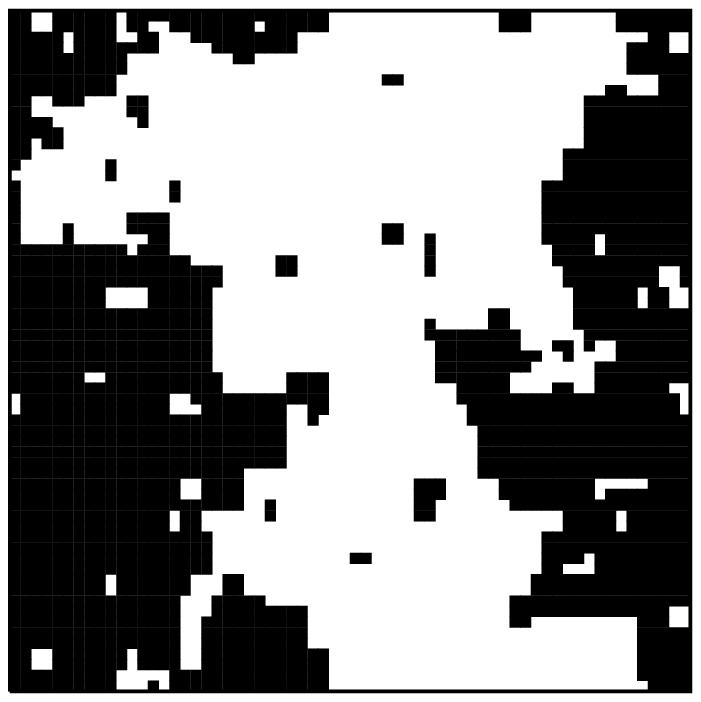,width=3.5cm}}
\centerline{(d)} } \caption{Snapshots of the configuration of the
staggered chirality in the fully frustrated array of size $L= 64$
at temperature $T$ and time $t$. (a) $T=0.1$ and $t=32$; (b)
$T=0.1$ and $t=1024$; (c) $T=0.4$ and $t=32$; (d) $T=0.4$ and
$t=1024$.} \label{fig:ffxy_conf}
\end{figure}

In order to help understand the temperature-dependence of the
dynamic exponent, we have also examined how ordering proceeds with 
time. Shown in Fig.~\ref{fig:ffxy_conf} are the snapshots at various 
time instants of ordering configurations in terms of the staggered
chirality. Figure~\ref{fig:ffxy_conf} (a) and (b) exhibit the
ordering process for quenching to low temperatures ($T=0.1$) while
(c) and (d) to higher temperatures ($T=0.4$), where black and
white regions represent domains with opposite (staggered)
chirality ordering.  One can observe distinctive features in the domain 
growth morphology in the two cases: 
At low temperatures, typical separation between consecutive corner
defects along a domain wall grows larger in time.
At higher temperatures, 
in contrast, the typical separation between corners remains more or 
less constant throughout the ordering process. 
Slow ordering at low temperatures ($T=0.1$) compared with
that at high temperatures ($T=0.4$) can also be observed here, in
agreement with the apparent temperature-dependence of $1/z_{q}$ 
obtained from the scaling analysis of the chirality correlation 
function. 

\begin{figure*}
\parbox{0.45\textwidth}{
\centerline{\epsfig{file=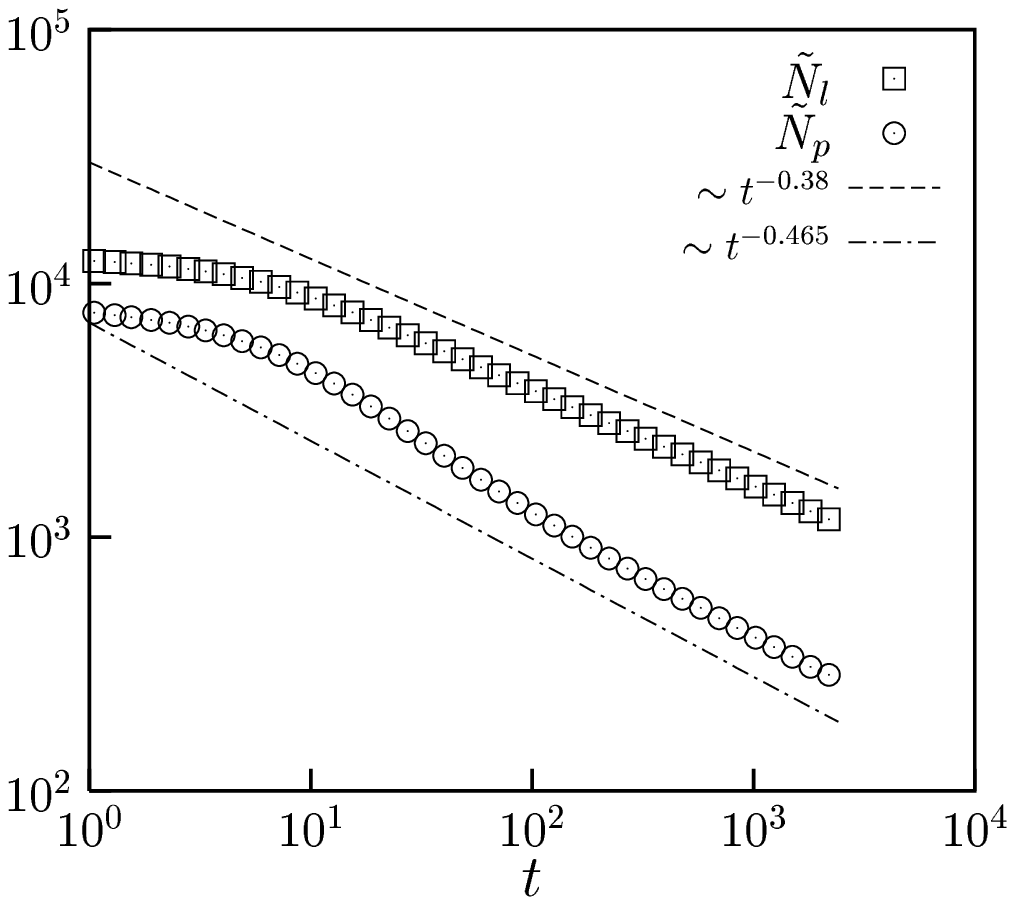,width=7cm}} \centerline{(a)}
\vspace*{0.5cm} \centerline{\epsfig{file=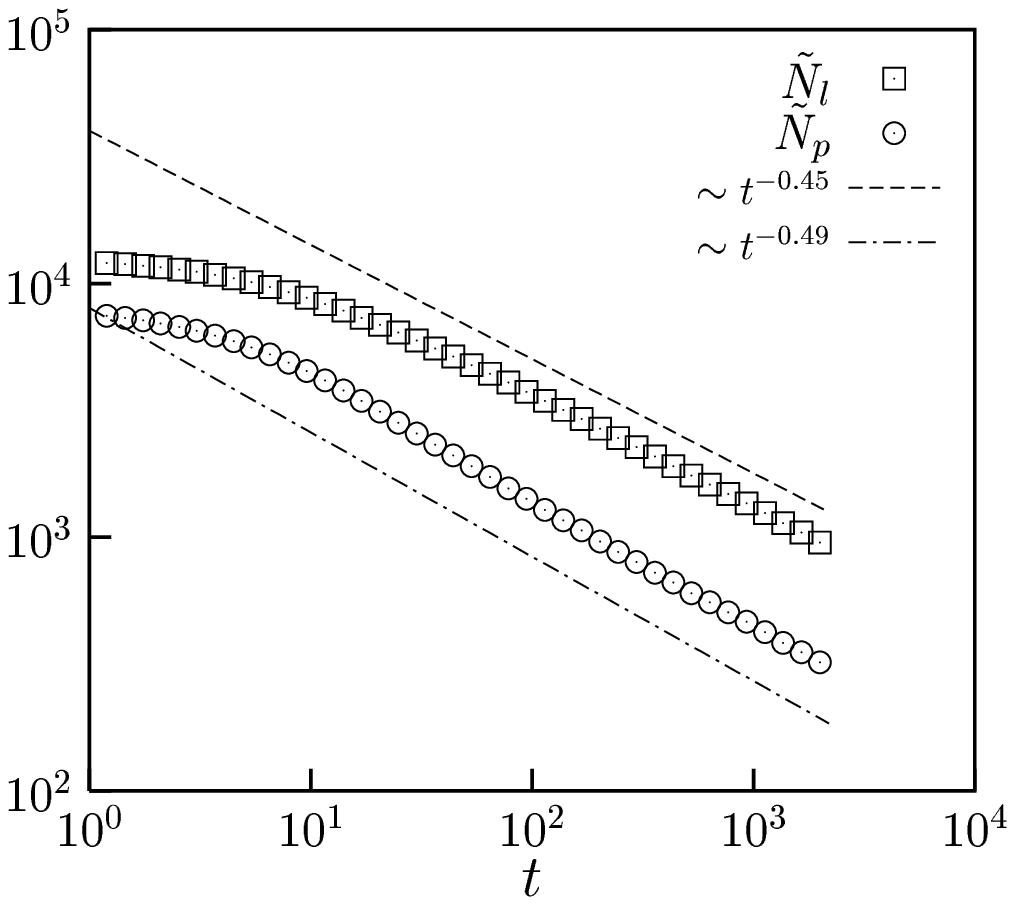,width=7cm}}
\centerline{(b)} } \hspace{0.05\textwidth}
\parbox{0.45\textwidth}{
\centerline{\epsfig{file=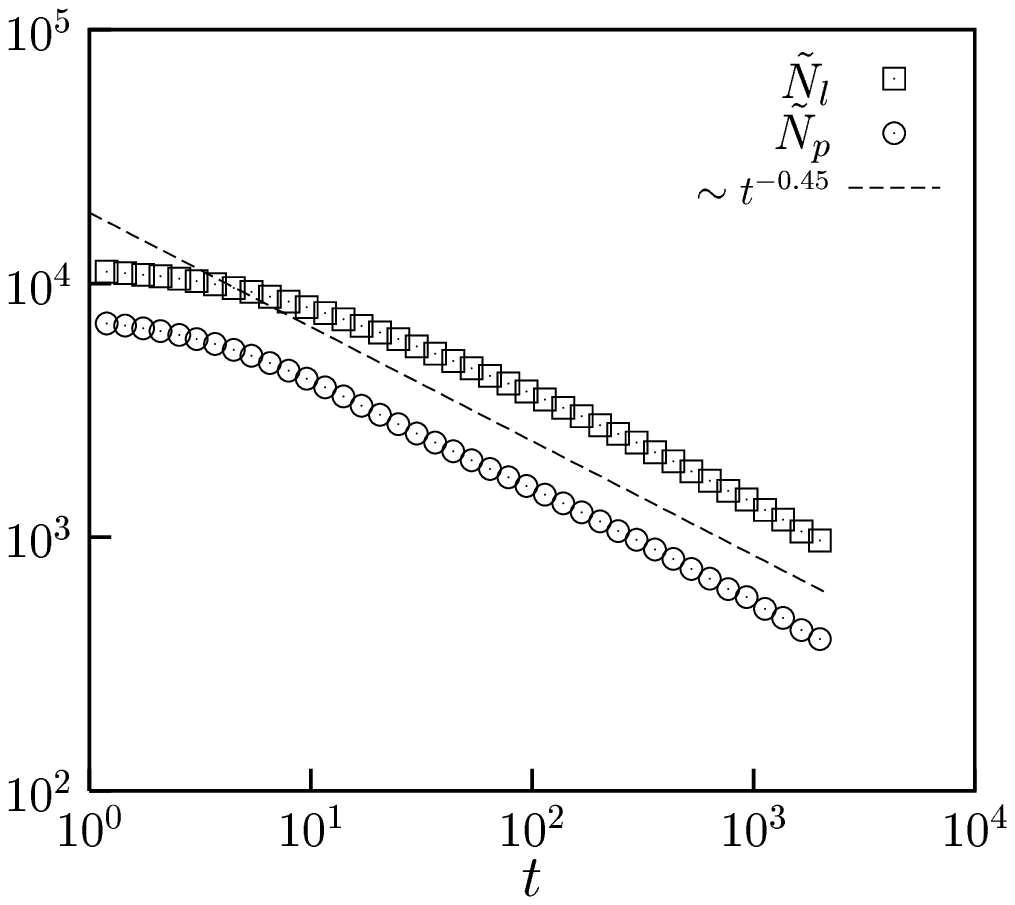,width=7cm}} \centerline{(c)}
\vspace*{0.5cm} \centerline{\epsfig{file=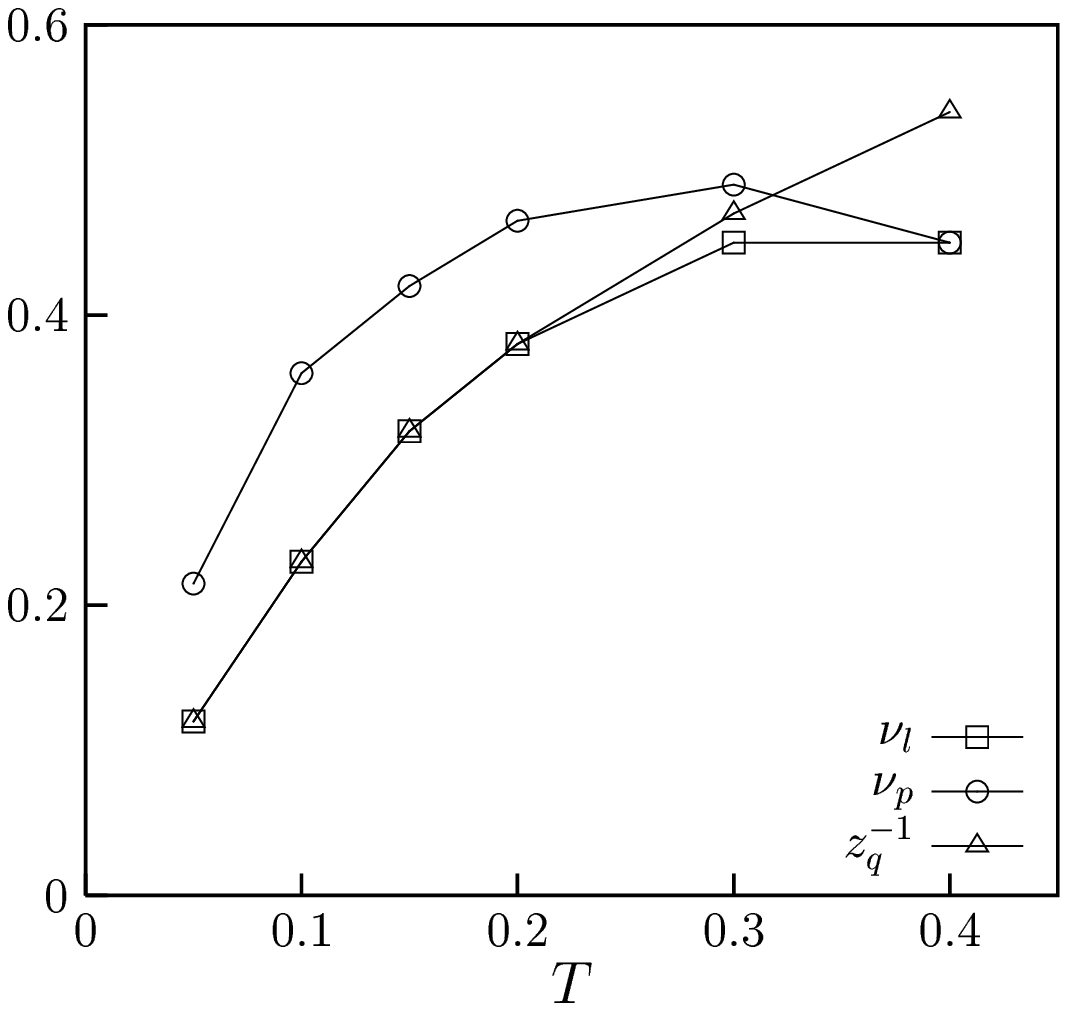,width=6.7cm}}
\centerline{(d)} } \caption{Relaxation of the excess length of the
line defects and of the excess number of point defects associated
with staggered chirality domains in the array of size $L = 128$ at
temperature (a) $T=0.20$; (b) $T=0.30$; (c) $T=0.40$.  Shown in
(d) is the temperature-dependence of the decay exponents $\nu_l$
and $\nu_p$ for line defects and for point defects, respectively,
together with the chirality growth exponent $1/z_q$.}
\label{fig:ffxy_defect}
\end{figure*}

The above morphological characteristics are related to the time
dependence of the number $N_{p}$ of point defects and of the total
length $N_l$ of line defects.  Simulations reveal power-law
relaxation toward equilibrium with temperature-dependent
exponents, as shown Fig.~\ref{fig:ffxy_defect}(a) to (c).  Namely,
we have $\tilde{N}_p (t) \equiv N_p (t) - N_p (\infty) \sim
t^{-\nu_{p}}$ and $\tilde{N}_l (t) \equiv N_{l}(t) -N_{l}(\infty)
\sim t^{-\nu_{l}}$, where $N_{p}(\infty)$ and $N_{l}(\infty)$ are
the values in equilibrium.  These equilibrium values, which depend
on the temperature, can easily be obtained from dynamic
simulations, beginning with one of the the known ground state
configurations.  The resulting equilibrium values are negligibly
small at low temperatures ($T \lesssim 0.2$) and the total amounts
$N_p$ and $N_l$ as well as the excess amounts $\tilde{N}_p$ and
$\tilde{N}_l$ display power-law decay.  The corresponding decay
exponents together with $1/z_{q}$ are shown in
Fig.~\ref{fig:ffxy_defect}(d) as functions of the temperature.
At low temperatures, we have $\nu_{p}> \nu_{l}$ and point defects 
decay away faster than line defects, resulting in growing separation 
between corner defects along domain walls.
At higher temperatures, on the other hand, we have 
$\nu_{p} \approx \nu_{l}$, and the average distance
between neighboring point defects along a line defect does not
change appreciably with time, similarly to the previous works\cite{sjl_ffxy,f2_cg}, 
which argued, based on such numerical results, that there exists a roughening 
transition at a finite temperature. 
However, it was argued in a recent work that due to the finite
excitation energy for the double-step kinks, the domain walls 
should be rough at any finite temperature.\cite{korshu_rough} 
If this is indeed the case, 
asymptotic decay exponents for the domain walls and corner charges
should be the same at all finite temperatures. 
Unfortunately, numerical confirmation of this would be very formidable 
due to the slow domain growth at low temperatures.

\section{Summary}

We have studied the non-equilibrium relaxation and coarsening
dynamics of two-dimensional superconducting arrays in the
overdamped limit under zero external driving current.  In the case
of unfrustrated arrays, we find that the friction constant of a
vortex remains finite in the limit of large extent of the vortex;
this is argued to lead to the absence of logarithmic factors in
the length scale and in the vortex number decay, which is in
reasonable agreement with simulation results. In the case of fully
frustrated arrays, relaxation and coarsening dynamics are
characterized by decay of line defects (chirality domain walls) as
well as of point defects (corner charges). Here strong
temperature dependence is found in the characteristics of the
late-time domain growth, which can be explained by activated
domain wall dynamics across the energy barriers with logarithmic size
dependence. 
In the early-time regime, on the other handr, there exists a relatively long 
period of transient dynamics which is independent of the temperature;
this may be attributed to the effects of initial random distributions
of the strong Lorentz force, making the thermal noise irrelevant.  
It would be of interest to investigate the coarsening dynamics of
Josephson-junction arrays under external magnetic fields and to compare with 
our simulation results.

\acknowledgments

We thank Bongsoo Kim for helpful discussions.  
This work was supported in part by 
the Korea Research Foundation through Grant No. 2000-015-DP0138.

\end{document}